\documentclass[fleqn,usenatbib]{mnras}

\usepackage{newtxtext,newtxmath}

\usepackage[T1]{fontenc}
\usepackage{ae,aecompl}

\pdfoutput=1
\usepackage{graphicx}	
\usepackage{amssymb}	

\title[Interior structure of the planetary bodies]{Interior structure of Mars and other rock-and-iron planetary bodies}

\author[A. Aitta]{
A. Aitta \thanks{E-mail: a.aitta@damtp.cam.ac.uk}
\\
Department of Applied Mathematics and Theoretical Physics,
Centre of Mathematical Sciences, University of Cambridge \\
Wilberforce Road, Cambridge, CB3 0WA, UK}
\pubyear{2018}

\makeatletter
\def\@journal{}
\makeatother

\makeatletter
\def\@journal{}
\def\@oddfoot{}
\def\@evenfoot{}
\makeatother

\begin{document}
\label{firstpage}
\pagerange{\pageref{firstpage}--\pageref{lastpage}}
\maketitle

\begin{abstract}
Here it is shown how to find the interior structure of a variety of rock-and-iron planetary bodies by using the rock density and some aspects of the core density as known for the Earth and using a convection principle for the iron-rich core. Convection minimizes both the density and temperature gradients inside the core fluid. This is achieved if the density of the core fluid is close to pure iron melting density at the core-mantle boundary, and the density has the smallest value possible for iron-rich melt at the inner core boundary. The critical iron densities for both pure iron and iron with  maximal light impurities were previously obtained utilizing Landau's theory of first order phase transitions with the most reliable experimental scaling.   The planetary interior density is found by iteratively calculating the gravity and pressure in small radial steps. Moment of inertia factors are also calculated and  agree well for the bodies for which we have accurate measurements: Moon, Mars and Mercury. Calculations are also made for the exoplanets Kepler-78b, K2-229b and Kepler-10b. All show iron-rich liquid in their cores. The lighter, solar objects are without an inner core, but the heaviest two exoplanets have a pure iron innermost core inside the inner core. The parameter range for a  growing inner core in the radius-mass plane is calculated. The magnetic field following the growth of an inner core protects life on Earth, and similarly for exoplanets. This guides the selection of exoplanets to study in the search for life. 
 
\end{abstract}


\section{Introduction}

 The layering in the rock-and-iron planets and  moons is due to the partial separation of the light rocky material  (mainly magnesium silicates) into the outer layers, and the heavy mainly iron-nickel material into the core. All terrestrial planets and the Moon are known to have some melt in their cores, but only the Earth is known, thanks to seismic studies \citep{lehmann_1936, dziewonski_1981}, to have a solidifying inner core which is generating  the Earth's magnetic field.  For any planetary body whose mass and radius are known, one can easily find the mean density of the body, but this can not inform one about the structure of the body  if one does not know more, for instance, the density of the materials it consists of, with estimates of the thickness of the layers. Particularly, if the body has a substantial amount of gas or ice on top, it may not be straightforward to conclude its structure deeper down. In this paper, the only planetary bodies analyzed are made of rock with iron-rich core. Here I solve in a systematic way the planetary interior structure for various planetary bodies having rocky mantle and at least partially molten and convecting iron-rich core, like all terrestrial planets in our solar system. I use the consequences of fluid dynamics and Landau's theory of  first order phase transitions  to estimate the densities in the cores, with the help of the Preliminary Reference Earth Model (PREM, \cite{dziewonski_1981}) particularly in the rock. This method has been used previously by \cite{aitta_2012} for Venus. Here this same strategy is applied also for the Moon, Mars, Mercury and the exoplanets Kepler-78b, K2-229b and Kepler-10b. All have iron-rich liquid in their cores. The lighter bodies are without an inner core, but the two heaviest have a pure iron innermost core inside an inner core made of iron with impurities.

\section{Methods}

\subsection{Consequences of core convection}

It is important to note that the material separation due to buoyancy forces is continuing inside a liquid core: convection tends to minimize both temperature and density gradients inside the fluid \citep{aitta_2012}. Since there is a big pressure range inside planetary cores, even a material of constant composition would have its density increase with pressure. Thus  core fluid containing some lighter components mixed with iron and nickel, would have a smaller density gradient  when such  fluid near the core-mantle boundary (CMB) is as dense as possible, that is, approximately corresponding to pure iron. As a consequence, the lighter material concentration would gradually increase with pressure, downward from the CMB. Similarly, the temperature gradient is the smallest when the core fluid at the CMB is as hot as possible. Thus at the CMB the core melt density and temperature are approximately those for pure iron at its melting temperature. Deeper down, both temperature and density increase, and the amount of light impurities  increases too. Still, both fluid density and temperature are increasing with pressure. The presence of a concentration gradient is against the common old belief that a convective core is compositionally uniform \citep{stevenson_1981}. 
  
\subsection{Landau's theory of first order phase transitions: Critical temperatures,  densities and concentrations}
\label{sec:maths} 

In order to understand a planetary core one needs to consider the properties of iron-rich material at high pressures and  temperatures close to the iron melting temperature. A very useful tool is  Landau's theory of first order phase transitions (\citet{landau_1937}, reprinted in \citet{ter_1965}, p. 193). It is based on the concept of an order parameter which is zero for an ordered phase like a solid and nonzero for a molten phase. The transition from liquid to solid is discontinuous in  entropy and density (or volume), and thus its mathematical description needs terms up to the sixth order in the order parameter in the Landau potential, which includes only even terms for symmetry reasons and adjustable coefficients in front  of the second and fourth order terms. A special point is where both these coefficients are simultaneously zero; this is called a tricritical point. A sixth order potential has at most three minima and thus this theory reveals that there are three critical conditions for the coexisting solid and  liquid states of a pure material \citep{aitta_2010a}. The most commonly known condition is the melting temperature which is the same as the solidification temperature. The theory gives also a condition how much the material temperature can be undercooled so that the liquid still exists but only as a metastable state, or how much the solid can be overheated as a metastable state. Instead of using temperature one can as well describe the phenomena using the material density for solid and melt at the melting temperature. These critical curves need to be calibrated  using the best available data. This was done by \citet{aitta_2006} for melting temperature and \citet{aitta_2010b} for melting density. Further experiments have not shown any need to change this calibration  \citep {dorogokupets_2017, sakaiya_2014}. However,  accurate very high pressure and temperature measurements are still lacking. This leads to some uncertainty at the highest pressures, especially close to or beyond the pressure at the tricritical point, about 800 GPa, relevant for many exoplanets.

The range of the metastable regime is widest when the pressure is small but it vanishes at  high pressure, since all three critical curves meet at the tricritical point. There the first order transition changes to be a continuous, second order phase transition, but this happens at such a high pressure and temperature (around 800 GPa, 8600 K) that the transition is suggested to be \citep{aitta_2006} from a solid to a plasma instead of to a liquid; however, this is still without experimental verification. At the tricritical point,  no change in  the density occurs, and there is no  latent heat. Only pressures inside the cores of K2-229b and Kepler-10b  exceed the tricritical pressure. 

The theory of tricritical phenomena predicts that the equation for the iron
melting temperature $T_M$ as a function of pressure $P$ is
\begin{equation}
    T_M(P)=T_{tc}-(T_{tc}-T_0)\left(\frac{P}{P_{tc}}-1\right)^2,\, \, \textrm{for } P<P_{tc},
    \label{eq:Tm}
\end{equation}
where $T_0$ is the iron melting temperature at normal (effectively zero) pressure
and ($P_{tc}$ , $T_{tc}$) is the tricritical point in the ($P$, $T$) plane. ($P_{tc}$ , $T_{tc}$ ) for iron was found using
the most reliable experimental melting data selected by \cite{aitta_2006}. 

This Landau theory also predicts \citep{aitta_2010a} that the lowest possible temperature for iron-rich melt, as a function of pressure, is
\begin{equation}
T_S(P)=T_{tc}-T_{tc}\left(\frac{P}{P_{tc}}-1\right)^2.
\label{eq:Ts}
\end{equation}
This gives the temperature at the ICB for any planet having a solidifying inner core.

 Temperature inside the outer core (OC) may be approximated by using  the same quadratic curve as the Earth's fluid core was found to have \citep{aitta_2012}:
\begin{equation}
T_{E,OC}(P)=2375.4+ 12.648 P - 7.9703 \cdot 10^{-3}P^2
\label{eq:Toc}
\end{equation}
where $P$ is in GPa, $T$ in K,
but shifting this down by a constant (or up for exoplanets) so that
$T_{OC}(P_{CMB}) = T_M(P_{CMB})$ for each planetary body .

The same theory predicts that the equation for the liquid iron density $\rho_M$ at its melting temperature as a function of pressure is
\begin{equation}
\rho_M(P)=\rho_{tc}-(\rho_{tc}-\rho_{0})\left(\frac{P}{P_{tc}}-1\right)^2 ,
\label{eq:rhom}
\end{equation}
where $\rho_0$ is the density at normal pressure and $\rho_{tc}$ the density at the tricritical point.
The expression for $\rho_{tc}$ is found \citep{aitta_2010b} using the known Earth's inner core boundary (ICB) density $\rho_{{melt}}^{{PREM}}$
and pressure $P_{ICB}$.

The lowest possible density for liquid iron or iron-rich melt, at $T = T_S$, is
\begin{equation}
\rho_S(P)=\rho_{tc}-(\rho_{tc}-\rho_{S0})\left(\frac{P}{P_{tc}}-1\right)^2 , 
\label{eq:rhos}
\end{equation}
where ${\rho_{S0}}$ is $\rho_S(P)$ at $P=0$.  $\rho_{S}(P)$ gives the ICB density for any planet having an iron-rich solidifying inner core. 

Density inside the outer core is set to have the same quadratic form \citep{aitta_2012} as in PREM:  
\begin{equation}
    \rho_{E,OC}(P)=7.8998 + 0.016413 P - 1.0343 \cdot 10^{-5} P^2\, ,
    \label{eq:rhoOC}
\end{equation} 
where $P$ is in GPa, $\rho$ in g/cm$^3$,
 but it is shifted down (or up for the exoplanets) so that $\rho_{OC}(P_{CMB}) = \rho_M(P_{CMB})$ for each planetary body.
 
 ICB is found where this outer core density equals to the critical density given by equation~(\ref{eq:rhos}). Inside the inner core (IC) the density is approximated to be the average of the densities given by  equations~(\ref{eq:rhom}) and~(\ref{eq:rhos}):
 \begin{equation}
\rho_{IC}(P)=\rho_{tc}-\left[\rho_{tc}-\frac{1}{2}(\rho_0+\rho_{S0})\right]\left(\frac{P}{P_{tc}}-1\right)^2 . 
\label{eq:rhoIC}
\end{equation}

Interestingly, there is a critical concentration of  lighter impurities that an  iron-rich melt can have  \citep{landau_2001, aitta_2010b}. If a particular planetary body has more, the excess amount of lighter material becomes bound into a solid phase with iron: thus an inner core solidification must happen. Planetary interior models having pure iron cores at pressures below $P_{TC}$ are without these important basics and cannot be very reliable. The maximum concentration ${c_i}$ of the light impurity as a function of pressure \citep{aitta_2010b} is
\begin{equation}
	c_i=1-c_{Fe}=1-\exp\left[{\frac{8L_0}{9 \mathcal{R} T_S}\left(\frac{P_{ICB}}{P_{tc}}-1\right)^3}\right]
	\label{eq:ci} 
\end{equation}
where the concentration $c_{Fe}$ includes all iron atoms together with the small amount of the very similar atoms such as nickel, likely to accompany iron in the planetary cores, ${L_0}$ is the latent heat of iron at normal pressure and $\mathcal{R}$  the gas constant.

\subsection{Planetary equations}

The mass $m(r)$  inside a sphere of radius $r$  is
\begin{equation}
	m(r) = 4\pi \int_{0}^{r} \tilde {r}^2 \rho(\tilde{r}) \, d\tilde{r} 
\end{equation}
 where $\rho$ is the density. For a body of radius $R$, the total mass is 
\begin{equation}
    M = m(R) = 4 \pi \int_{0}^{R} r^2 \rho(r) \, dr \, .
    \label{eq:totalmass}
\end{equation}
The pressure at radius $r$ and hence depth $h = R - r$  is 
\begin{equation}
    P(h) = \int_{0}^{h} g(\tilde{h}) \rho(\tilde{h}) \, d\tilde{h} 
	\label{eq:Ph}
\end{equation}
where the gravitational acceleration at depth $h$ is 
\begin{equation}
    g(\tilde{h})=\frac{G m(R-\tilde{h})}{{(R-\tilde{h})}^2} \, , 
\end{equation}
with $G$ Newton's constant. 

The mean moment of inertia $I$ is
\begin{equation}
    I = \frac{8\pi}{3} \int_{0}^{R} r^4 \rho(r) dr \, .
    \label{eq:moment of inertia}
\end{equation}  
The moment of inertia factor $MOI$ for a planetary body is 
\begin{equation}
    MOI = \frac{I}{MR^2} \, .
    \label{eq:MOI}
\end{equation} 

\subsection{Density profile}

We know the Earth's rock density as a function of pressure $P$, and it can be well modeled \citep{aitta_2010b} with two layers: linear in $P$ in the upper mantle where $P$ is below 23.83 GPa 
\begin{equation}
    \rho(P) = 3.2872+0.032775 P 
    \label{eq:UM}
\end{equation} 
and quadratic in $P$ in the lower mantle where $P$ is above 23.83 GPa
\begin{equation}
    \rho(P) = 4.0578+0.014338 P - 2.3987 \cdot 10^{-5} P^2  \, ,
    \label{eq:LM}
\end{equation} 
where $P$ is in GPa, $\rho$ in g/cm$^3$.
All bodies here are assumed to have a similar rock density. A separate crust is employed only for the Moon.  

The temperature and density gradients in the core would become smallest when there is as dense and as hot material as possible at the smallest pressures occuring there, that is at the CMB, and as light and cool material as possible at the highest pressure, at the ICB or in the centre if there is no inner core. So the fluid density at CMB is as close as possible to the pure iron density given by equation~(\ref{eq:rhom}) and its temperature is approximately the  iron melting temperature from equation~(\ref{eq:Tm}). And the fluid density at ICB is as close as possible to the lower critical density curve given by equation~(\ref{eq:rhos}) and its temperature  is approximately  on the lower critical temperature curve from equation~(\ref{eq:Ts}). 

Fluid densities and temperatures between these extremes can be approximated by equation~(\ref{eq:rhoOC}) and equation~(\ref{eq:Toc}) with the appropriate shifts. 

After convection has rearranged the material this way there might be too much light matter at the highest pressures. The Landau theory tells us also what is the maximum concentration of the light impurities at each pressure through equation~(\ref{eq:ci}). At the Earth's ICB pressure this is 5.1 mol.\%, but only 3.8 mol.\% at the Earth's centre pressure. If the core fluid in the early Earth had more than this 3.8 mol.\% of light impurities, the fluid  would not be stable. The only option is for the core fluid  to reduce its pressure because at smaller ICB pressures there would be space for more impurities according to equation~(\ref{eq:ci}). To do this it needs to solidify the core fluid at the highest pressures by making a solid inner core,  thus making the fluid container smaller. We can conclude from the PREM densities that about a half of the light material forms an alloy with iron in the solid inner core but the rest is released into the core fluid. This release of light material is believed to drive the Earth's magnetic field. The inner core growth would continue only as long as there is too much light material in the deepest part of the fluid.  

The Earth's inner core is not pure solid iron, but its  density  is approximately half-way between  pure iron density $\rho_M$ and the density of the least dense iron-rich melt $\rho_S$. This feature is assumed here for all bodies with an inner core using the equation~(\ref{eq:rhoIC}), because we do not know any better. 

According to the Landau theory there is no melt beyond the tricritical pressure. As remarked earlier, there the solid might change to a warm dense plasma instead of a  liquid, but this is still unconfirmed experimentally.
Thus for the two most massive objects studied, the exoplanets K2-229b and  Kepler-10b, the innermost cores have pressures beyond about 800 GPa and are thus without light impurities. They are modeled to be full solid  without any density change with increasing pressure:
\begin{equation}
	\rho(P) = 16.14 \,\, \rm{ g/cm^3 } \, \, \textrm{ for }  $$P \geq P_{tc}$$.
    \label{eq:tc}
\end{equation} 
We do not have any experiments to show whether a density increase with pressure takes place in reality.

\section{Results}

The schedule is the same for each planetary body. All cores are iron-rich and molten near the CMB. The radius of the CMB is not known beforehand, but at the CMB the fluid is postulated to have the iron melting temperature and the pure iron  density at its melting temperature. This was found to be approximately true earlier while studying the Earth \citep{aitta_2010b} and it helped to understand that  the  convection in the fluid core would minimize both density and temperature gradients. This challenges a common suggestion in the present literature that at the CMB the fluid has the most impurities \citep{helffrich_2010}, or it is homogeneous inside the core.

An elementary numerical integration designed to find $R_{CMB}$ to an accuracy of 1 km was employed. After an initial guess for the radial pressure profile, the density $\rho(P)$ is calculated for each zone (upper mantle ~(\ref{eq:UM}), lower mantle ~(\ref{eq:LM} with an appropriate shift), outer core ~(\ref{eq:rhoOC}) and inner core ~(\ref{eq:rhoIC}) and the innermost core if relevant ~(\ref{eq:tc})) and the mass $m(r)$ is then integrated in small radial steps, and the gravitational acceleration $g(r)$ and moment of inertia factor are then calculated. The location of the CMB is first guessed and  $\rho_{CMB}$  is set to be $\rho_M$ from equation~(\ref{eq:rhom}). 
The total planetary mass integral $M = m(R)$ is found and used to adjust the location of the CMB to produce $M$ in a reasonable range. Then a new pressure profile is calculated from the density profile and corresponding $g$ using equation~(\ref{eq:Ph}). The process is then iteratively repeated until the mass is as correct as possible with a stabilized pressure profile. For the most massive objects, densities beyond the tricritical pressure are needed, but they are just estimated to be constant from equation~(\ref{eq:tc}) as in the theoretical model by \citet{aitta_2006, aitta_2010a}, since we do not have accurate experimental data to know any better.

\subsection{Moon}

\begin{figure}
	\includegraphics[width=0.5 \textwidth]{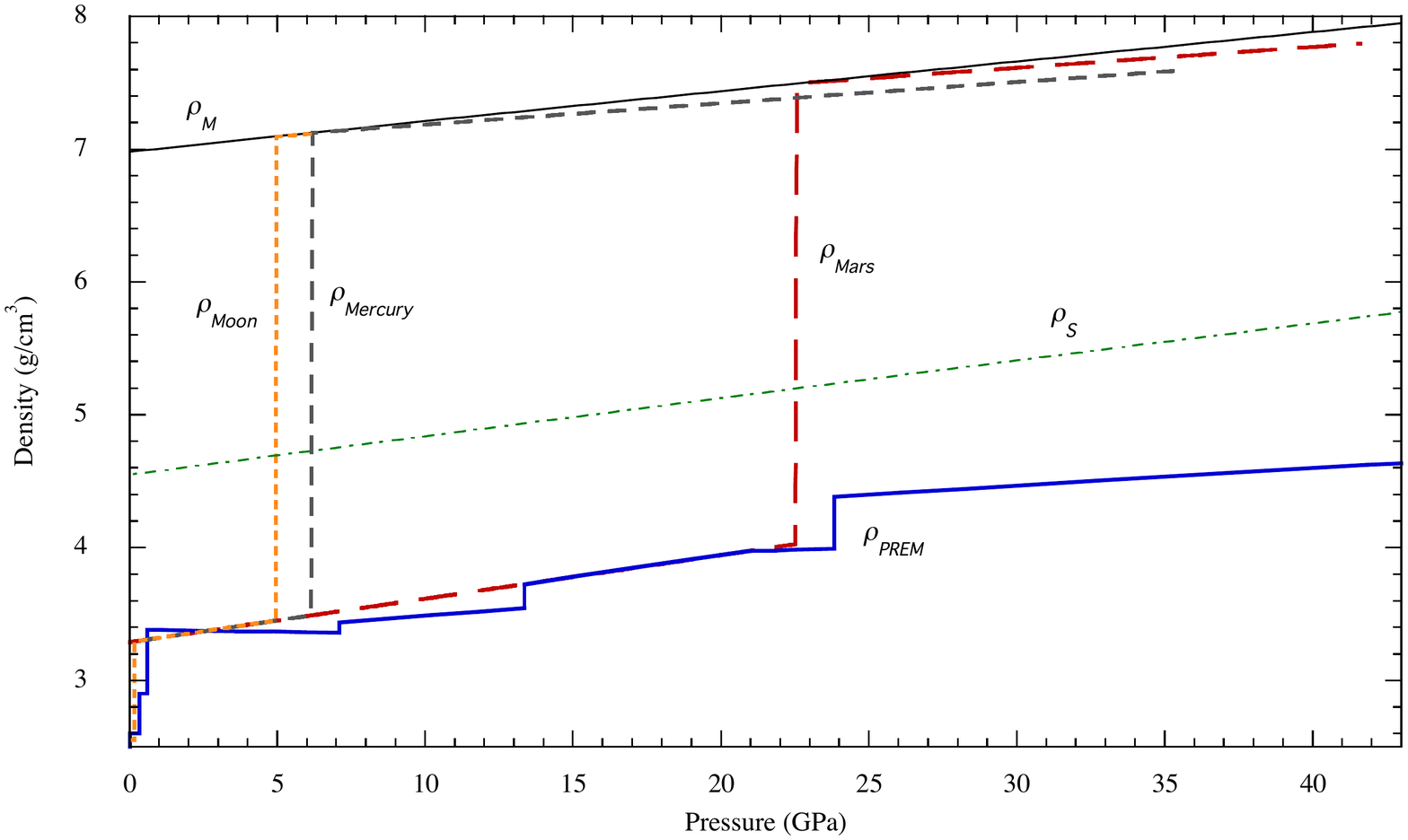}
    \caption{Density profiles as a function of pressure for Moon, Mercury and Mars. Also iron density $\rho_M$ at its melting temperature is shown, together with the lowest density possible for an iron-rich melt $\rho_S$ and PREM density profile $\rho_{PREM}$ for the Earth.}
    \label{fig:MaMeMo4_figure}
\end{figure}

A separate crust density 2550 kg/{m$^3$} and thickness 38 km \citep{wieczorek_2013} is assumed only for the Moon. Modeling the crust, upper mantle and core one finds by iteratively integrating that $M = 7.3473 \cdot 10^{22}$ kg, very close to the known value $M = 7.3477 (\pm 0.0033) \cdot 10^{22}$ kg. Moon's core radius is found to be $R_{CMB} = 403$ km and the moment of inertia factor $MOI = 0.38889$, rather close to the estimate $0.3935 \pm 0.0002$ (\citet{konopliv_1998} but here for the mean radius). In addition, one finds pressures $P_{centre}$ = 6.11 GPa at the centre and $P_{CMB}$ = 4.95 GPa, and temperature $T_{CMB}$=1896 K from equation~(\ref{eq:Tm}).
One obtains the estimate  $T_{centre}$=1910 K at the centre by using  equation~(\ref{eq:Toc}) with the shift for Moon. The core/mass ratio is  0.0266 and is similar to \citet{stacey_2005}: 0.024$\pm$ 0.002. These results are all listed in Table~\ref{tab:example_table}.   
The density profile is presented in Fig.~\ref{fig:MaMeMo4_figure} together with the results for  Mars and Mercury which are discussed next. 

\subsection{Mars}

Mars' radius is $R$ = 3390 km and mass is $M = 6.4185 \cdot 10^{23}$ kg. Mars' mean moment of inertia factor is rather well-known to be $0.3645 \pm 0.0005$ \citep{rivoldini_2011}, but we do not know its crust density or the crust thickness. While  trying to include a crust  in these calculations I was not able to get a good value for $MOI$.
However, by having upper mantle density  decreasing  linearly to the surface, probably a consequence of  the extensive resurfacing by volcanism over Mars' history, one eliminates the unknown thickness and density of its crust, and obtains an accurate value for $MOI$.
By iterative integration, one finds  $M = 6.4185 \cdot 10^{23}$ kg as expected and  $MOI$=0.36347. This linear density profile for the crust is an interesting finding indicating lesser crust differentiation. It is applied next to  Mercury, since its crust density and thickness are unknown, too, and neither planets have plate tectonics, the main process in Earth differentiating the crust from the mantle. The fact that Moon has a clear crust, even though it does not have plate tectonics either, can follow from its different birth mechanism due to the ejection from the Earth.  One obtains for Mars that $R_{core}$ = 1522 km, $P_{CMB}$=22.5 GPa, $P_{centre}$=41.63 GPa. Its core size is  in the range  1520-1840 km \citep{yoder_2003}. Both pressures agree well with a recent range in \citet{pommier_2018}. The core/mass ratio is  0.175 and is rather close to \citet{stacey_2005}: 0.156$\pm$ 0.010.  Results are listed in Table~\ref{tab:example_table}. The density profile is presented in Fig.~\ref{fig:MaMeMo4_figure}. 

\subsection{Mercury}

A linear decrease of the upper mantle rock density to the surface is again employed. 
Using Mercury's radius $R=2439$ km  one obtains $M = 3.3022 \cdot 10^{23}$ kg and $MOI$  = 0.34516 as expected: compare $0.346 \pm 0.014$ in \citet{rivoldini_2013}. $R_{CMB}$ = 1965 km is just on the lower boundary of a result by \citet{rivoldini_2013}. In additition, $P_{centre}$ = 35.77 GPa, $P_{CMB}$=6.16 GPa, $T_{CMB}$=1917 K, $T_{centre}$=2281 K and the core/mass ratio is 0.704. This ratio is close to \citet{stacey_2005}: 0.679 $\pm$ 0.015. These results are listed in Table~\ref{tab:example_table}. 
The density profile is presented in Fig.~\ref{fig:MaMeMo4_figure} along with the results for Moon and Mars.

\subsection{Kepler-78b}
	
\begin{figure}
    \hspace*{-2.5cm}
	\includegraphics[scale=0.45]{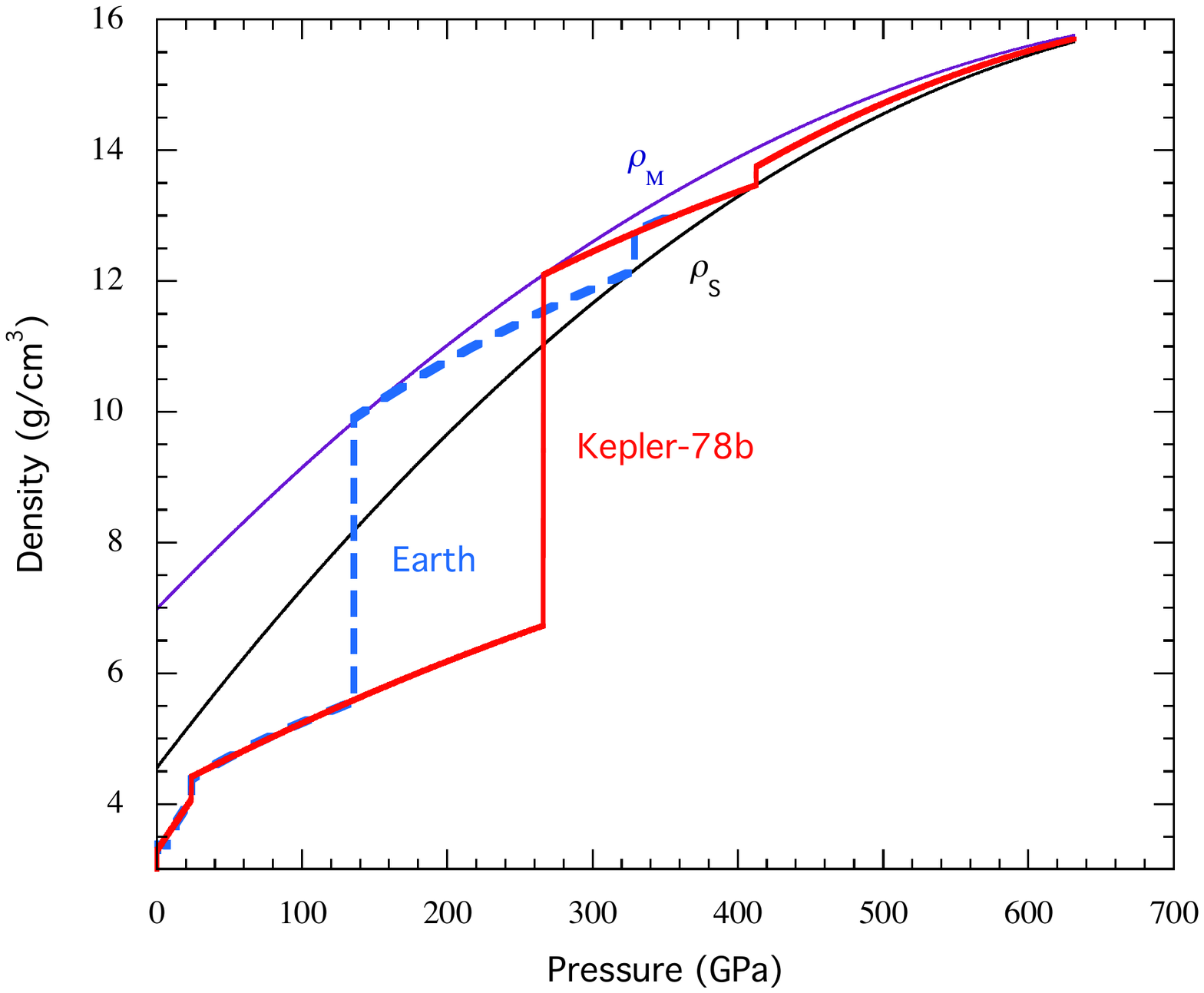}
    \caption{Density profile as a function of pressure for Kepler-78b.  Densities $\rho_M$, $\rho_S$ and PREM Earth density are as in Figure ~\ref{fig:MaMeMo4_figure}.}
    \label{fig:K78bav_figure}
\end{figure}

Recently two groups \citep{howard_2013, pepe_2013}  independently published their findings  for the exoplanet Kepler-78b, giving the planet's mass and radius with a reasonable accuracy. This planet was concluded to be rocky like the Earth, not very much larger but with much hotter temperature due to the closeness of its star.  The relative similarity in the size and  mass to the Earth allows one to analyze the planet's internal structure further, including the presence of a liquid OC and solidifying IC using a technique employed earlier for  Venus. A convecting OC would generate a magnetic field which may be possible to observe in the future.
The average mass and radius obtained in those two studies are $1.775 M_E$ and $1.1865 R_E$ (expressed using Earth's mass and radius, respectively)  corresponding to about $10.601 \cdot {10}^{24}$ kg and 7560 km, which have been used in this study. 
Kepler-78b is Earth-like enough to consider it to have an upper mantle, a lower mantle, and a liquid OC and solidifying IC.  For this averaged planet's $M$ and $R$ one can calculate its pressure profile and various radii, pressures and temperatures with the core fraction and $MOI$. 
These are presented in Table~\ref{tab:example_table}. 
The core mass fraction is 0.26,  belonging to the range $0.32 \pm 0.26$ of \citet{grunblatt_2015}. 
The figure ~\ref{fig:K78bav_figure} shows Kepler-78b's density as a function of pressure together with the Earth's PREM density. One can see that the IC is much more substantial for this exoplanet. But qualitatively both planets have a similar structure.

The integration gives  $R_{CMB}$=3572 km and $P_{CMB}$=266 GPa; the IC radius is 2627 km with $P_{ICB}$=413 GPa and $P_{centre}$=631 GPa.

\subsection{K2-229b}

This metal-rich exoplanet has mass $M=(2.59 \pm 0.43) M_E$ and radius $R=(1.165 \pm 0.066) R_E$ \citep{santerne_2018}. 

\begin{figure}
\hspace*{-2.5cm}
\includegraphics[scale=0.45]{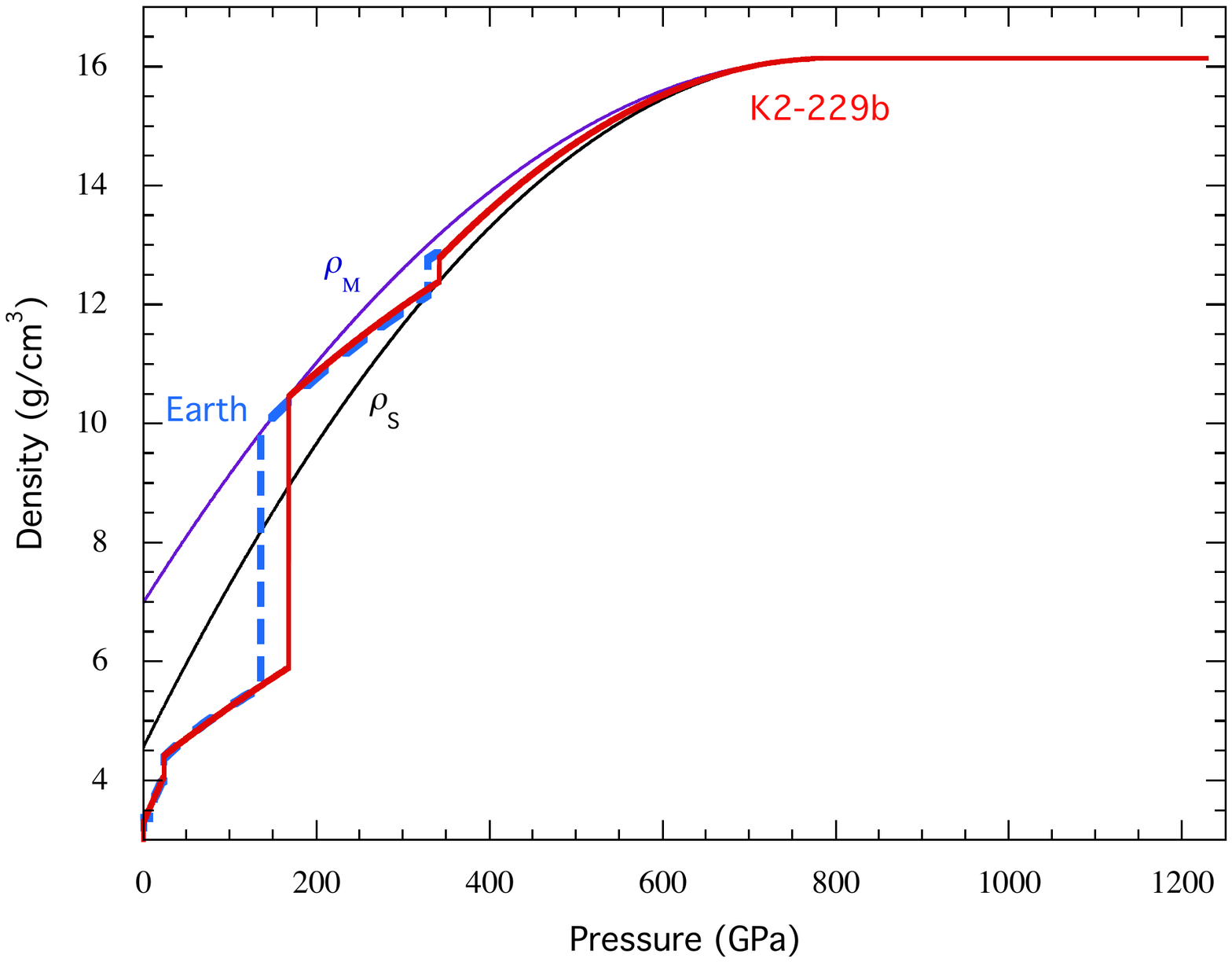}
	\caption{Density profile as a function of pressure for K2-229b. $\rho_M$, $\rho_S$ and PREM Earth density are as in Figure ~\ref{fig:MaMeMo4_figure}.}
    \label{fig:K2229bL_figure}
\end{figure}

The integration gives  $R_{CMB}$=5741 km and $P_{CMB}$=168 GPa. Figure ~\ref{fig:K2229bL_figure} shows the density of K2-229b as a function of pressure and other results are in Table~\ref{tab:example_table}.  The IC size is 5050 km with $P_{ICB}$=342 GPa and the size of the innermost IC (which  in this model is made of pure iron with a constant density) is 3454 km reaching to the pressure $P_{TC}$=793 GPa from $P_{centre}$=1228 GPa. Its inner structure resembles an Earth-like planet. Its outer core is comparable to the Earth, but with a much more massive inner core, a big part of it being pure solid iron. Its core mass fraction 0.72 is Mercury-like \citep{santerne_2018} but the layering is more complicated:  mantle and core  have two and three layers, respectively, while Mercury has only one of each.

\subsection{Kepler-10b}

\begin{figure}
\hspace*{-2.5cm}
\includegraphics[scale=0.45]{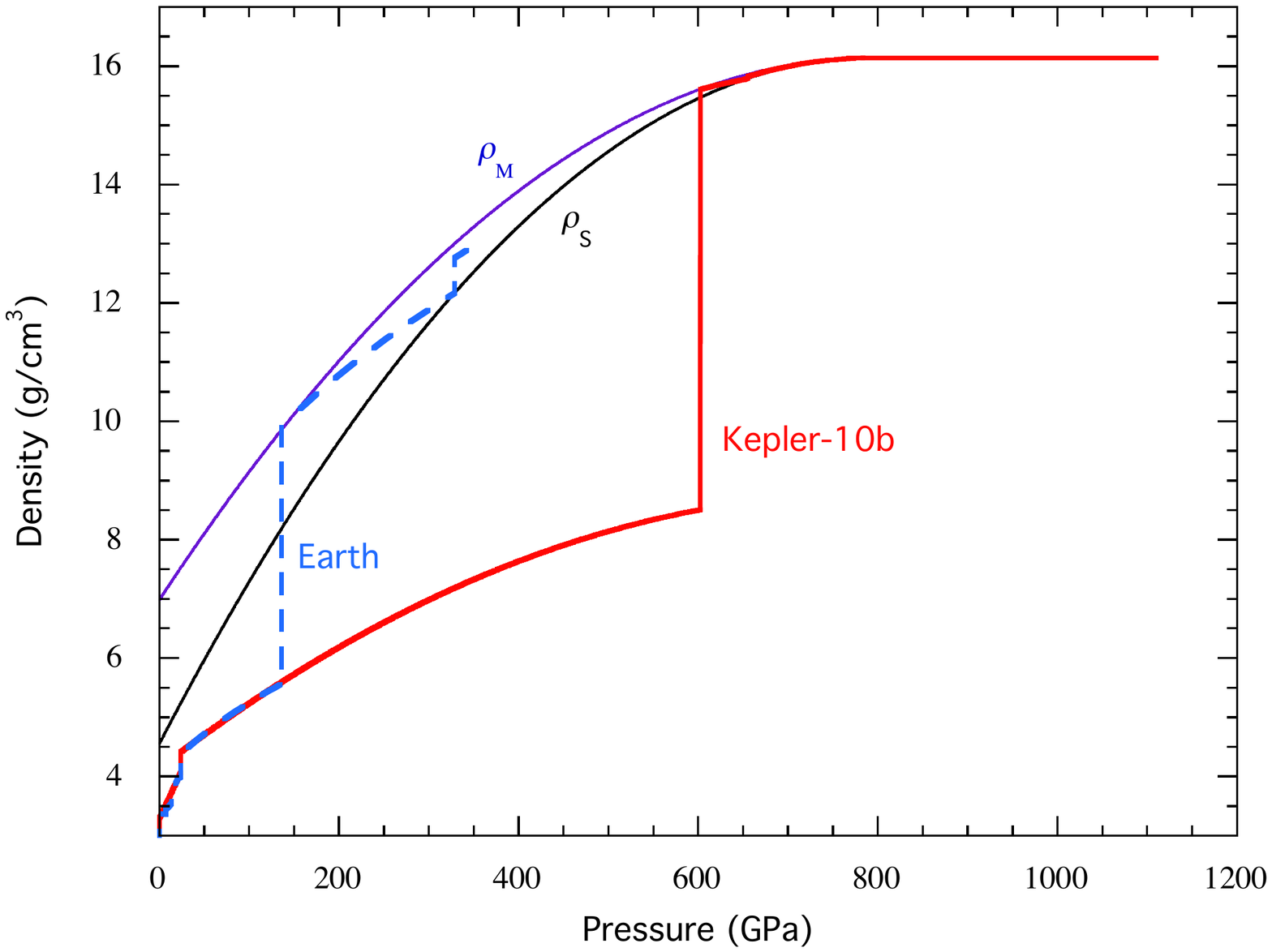}
	\caption{Density profile as a function of pressure for Kepler-10b. $\rho_M$, $\rho_S$ and PREM Earth density are as in Figure ~\ref{fig:MaMeMo4_figure}.}
    \label{fig:K10b_figure}
\end{figure}

Kepler-10b is expected to be rocky consistent with an Earth-like composition \citep{seager_2007}. Its currently  most accurate mass is $(3.72 \pm 0.42) M_E$ \citep{weiss_2016} and the radius $R=(1.47 \pm 0.02) R_E$. These results are shown in Fig.~\ref{fig:K10b_figure} and in Table~\ref{tab:example_table}. 
This analysis shows Kepler-10b has a very substantial mantle, and thus its core is at very high pressures: most of it is pure solid iron. That is why these results become sensitive to the uncertainties in this modeling. The mantle reaches over four times the pressures of the Earth's mantle and we do not know  mantle rock densities well at such high pressures. Most of the core is also  in the region without good experimental confirmation of the iron density.
However, the core mass fraction 0.158 compares very  well with $0.17 \pm 0.12$ based on both HARPS-N and HIRES \citep{weiss_2016}.
But very recently, \citet{wicks_2018}  has modeled this exoplanet by having 15 \% Si in its core and shows  $P_{centre}$ to be $\sim$  1130 GPa,  $\rho_{centre}$ $\sim$ 17 g/cm$^3$ and $R_{CMB}$ $\sim$ 4450 km. These compare reasonably well with the results here:
1110 GPa, 16.14 g/cm$^3$ and 3741 km.
But both results differ substantially from the early analysis by \citet{wagner_2012} giving correspondingly 2230 GPa, $\sim$ 21 g/cm$^3$ and 6070 km, which for the density alone  is close to \cite{wicks_2018}'s result for pure iron core:   1330 GPa, $\sim$ 21 g/cm$^3$ and 3823 km.

\begin{table*}
	\caption{Planetary radius and mass from literature. Results for various internal values of $R$, $P$, $T$, $M_{core}/M$ and $MOI$ together with the values for Earth radii from \citet{dziewonski_1981}, temperatures from \citet{aitta_2006} and Venus' core radius, $P$'s $, T$'s  and $MOI$ from \citet{aitta_2012}.}
	\label{tab:example_table}
	\begin{tabular}{lcccccccccccr} 
		\hline
		Planetary & $R$ & $M$ & $R_{CMB}$  & $P_{CMB}$ & $T_{CMB}$  & $R_{ICB}$  & $P_{ICB}$  & $T_{ICB}$ & $P_{centre}$ & $T_{centre}$ & $M_{core}/M$ & $MOI$ \\

         body & km & kg & km &  GPa &  K &  km & GPa & K &  GPa & K & &\\
         \hline
		Moon & 1737 & $7.3477 \cdot 10^{22}$ & 403 & 4.95 & 1896 &  & & & 6.11& 1910 & 0.0266 & 0.389 \\
		Mercury &2439 & $3.3022 \cdot 10^{23}$ & 1964 & 6.16 & 1917 & & & & 35.8 &2281 & 0.704  & 0.345\\
		Venus &6052 & $4.8680 \cdot 10^{23}$ & 3228 & 114 & 3630 & & & & 275 & 5200 & 0.289 & 0.338\\
        Earth & 6371 & $5.9736 \cdot 10^{24}$ & 3480 & 136 & 3945 & 1221 & 329 & 5670 & 364 & 6350 & 0.325  & 0.330\\
        Mars &3390 & $6.1485 \cdot 10^{23}$ & 1522 & 22.5 & 2192 & & & & 41.6 &2425& 0.175 & 0.363\\
        K-78b & $1.1865R_E$ &$1.775 M_E$ & 3572 & 266 & 6187 & 2631 & 413 & 6340 & 631 & 8300 & 0.26 & 0.332\\
        K2-229b & $1.165 R_E$ & $2.59 M_E$& 5741 & 168 & 4580 & 5050 & 342 & 5840 & 1228 & 8600 & 0.72 & 0.310\\ 
        K-10b & $1.47 R_E$& $3.72 M_E$ & 3741 & 603 & 8238 & 3535 & 656 & 8371 & 1110 & 8586 & 0.158 & 0.335\\
		\hline
	\end{tabular}
\end{table*}

\subsection{Inner core range}

\begin{figure}
\hspace*{-3.1cm}
\includegraphics[scale=0.475]{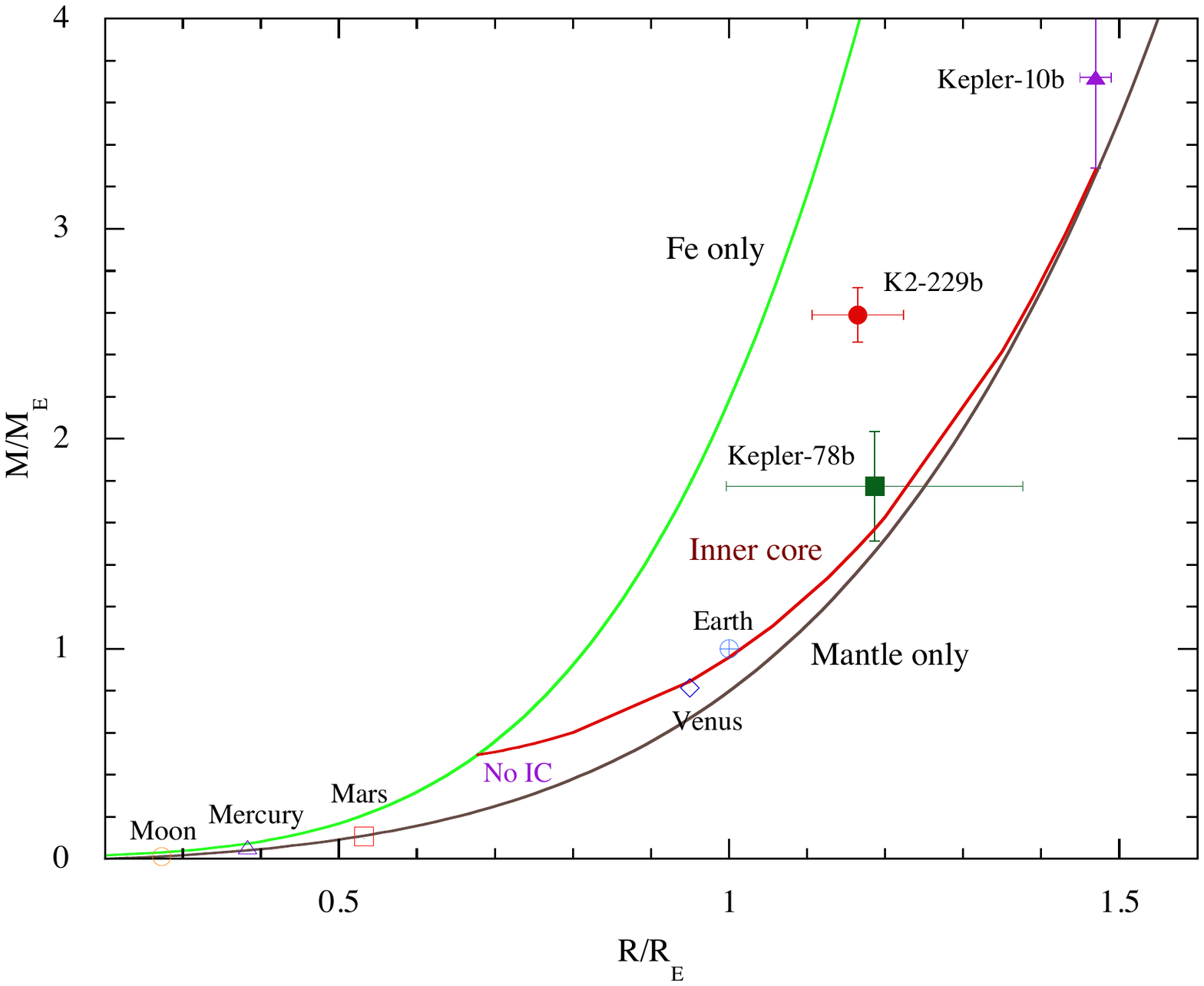}
    \caption{ The range in the radius-mass plane for a planet to have a solidifying inner core, a prerequisite for a magnetic field to protect life from stellar winds. The limiting curves for bodies with only a mantle and only pure iron are also shown as well as the data for the planetary bodies considered here, including  the Earth and Venus.}
    \label{fig:ICB_figure}
\end{figure}

Now one can find the critical conditions for the qualitative change in the internal structure. 
For the first time, the range in the radius-mass plane is calculated for any planet to have a solidifying inner core, a prerequisite for a magnetic field to protect life from stellar winds.
By using a selection of suitable radii one can find the limiting mass when a rock-iron planet can have an inner core. The criterion is that, at $P_{centre}$ its density is $\rho_S(P_{centre})$. This is shown in Fig.~\ref{fig:ICB_figure} by the red curve. The polynomial fit gives 
\begin{equation}
\begin{split}
\frac{M}{M_E}  =  1.2688-3.4195 \frac{R}{R_E} + 4.8299  \left(\frac{R}{R_E}\right)^2 \\
-3.1052  \left(\frac{R}{R_E}\right)^3+1.385  \left(\frac{R}{R_E}\right)^4. \\
\end{split}
\end{equation}
Above this curve, a planet has an inner core, like the Earth. Below this curve, the planet does not have an inner core, like Venus and the small terrestrial planets and the Moon. 

The lower curve (brown) shows planetary masses whose density never reaches the liquid iron melting density $\rho_M$ at the centre:
\begin{equation}
\begin{split}
\frac{M}{M_E}
 = 5.3985\cdot10^{-3}-0.16314 \frac{R}{R_E} + 0.81671 \left(\frac{R}{R_E}\right)^2 \\
-0.72195 \left(\frac{R}{R_E}\right)^3+0.86103 \left(\frac{R}{R_E}\right)^4. \\
\end{split}
\end{equation}
Such planets have mantle rock only (without a separate crust). 
On the upper curve (green) the planetary masses have $\rho = \rho_M$ at all pressures, modeling pure Fe at its melting temperature:

\begin{equation}
\begin{split}
\frac{M}{M_E}  = -5.8632 \cdot10^{-4}
+0.06742 \frac{R}{R_E}
+ 0.018619  \left(\frac{R}{R_E}\right)^2 \\
-1.7356 \cdot10^{-4}  \left(\frac{R}{R_E}\right)^3
+2.0951  \left(\frac{R}{R_E}\right)^4. \\
\end{split}
\end{equation}

\section{Conclusions}

The Landau theory of first order phase transitions is an excellent way to probe the planetary structure of rock-iron bodies in our solar system and beyond. By assuming the rock density and the quadratic behaviour  of core density as a function of pressure stay similar to the Earth one can calculate various quantities inside the planets.  By comparing them to the values obtained by others with different methods and assumptions one finds very close similarities to \citet{stacey_2005} for the terrestrial bodies.  For the first time, the range in the radius-mass plane is calculated for any planet to have a solidifying inner core, a prerequisite for a magnetic field to protect life from stellar winds.

\section*{Acknowledgements}

I thank Caroline Dorn for letting me know about the recent work of Santerne et al. (2018).
The results for Moon, Mercury, Mars and Kepler-78b were first presented (June 2014) in ``From tricritical phenomena describing symmetry breaking in fluid flow to tricritical phenomena in the iron melting curve: insights into planetary interior structure, core temperature and composition" at the International Conference on Phase Transitions at Low Temperatures, Pattern Formation and Turbulence in honor of the 80th birthday of Guenter Ahlers.

\bibliography{corestr} 

\bsp	
\label{lastpage}

\end{document}